\documentstyle[12pt]{article}
\def\beq{\begin{eqnarray}}
\def\neq{\end{eqnarray}}
\def\eq#1{eq.\ (\ref{#1})}
\def\eqs#1#2{eqs.\ (\ref{#1}, \ref{#2})}

\def\bitem{\vspace{-7pt}\bibitem}
\def\nl{\newline}
\def\slb{{\scriptstyle ( }}
\def\srb{{\scriptstyle ) }}

\def\dif{{\rm d}}
\def\rarrow{\rightarrow}

\def\etal{\mbox{\em et.\ al.}}

\def\I{\'\i{}}
\def\ao{\mbox{\~ao}}
\def\plb{Phys. Lett. B\ }
\def\npb{Nucl. Phys. B\ }

\def\zpc{Z. Phys. C\ }

\def\prd{Phys. Rev. D\ } 

\def\ODW{{\cal O}_{DW}}
\def\ODB{{\cal O}_{DB}}
\def\OF1{{\cal O}_{\Phi,1}}
\def\OBW{{\cal O}_{BW}}
\def\OW{{\cal O}_{W}}
\def\OB{{\cal O}_{B}}
\def\OWWW{{\cal O}_{WWW}}
\def\ODWW{{\cal O} '_{DW}}

\def\fDW{f'_{DW}}
\def\fWWW{f'_{WWW}}
\def\mfw{{{f'_{DW}\, g^2} \over {\Lambda ^2}}}
\def\mfb{{ f'_{DB}\, {g '}^2 \over {\Lambda ^2}}}
\def\kk{\left(k^2 \gamma^\mu - k^\mu \! {\not{\! k}} \right)}
\def\kks{k^\mu \!\! \not{\!  k}}

\def\ga{\gamma}
\def\Ga{\Gamma}
\def\DG{\Delta \Gamma}
\def\DGg{\Delta \Gamma_\gamma}

\def\mw{M_W}
\def\mz{M_Z}                

\def\g0{\Gamma_0}

\def\gf{{\rm G_\mu}}

\def\sm{{\rm SM}}

\def\da2{(\dif \theta^2 + \sin^2\theta \, \dif \phi^2)}
\def\dte2{\dif {\vec \theta}^2}
\def\dsi2{\dif \sigma^2}

\begin{document}
\baselineskip=20pt



\vspace{-60pt}
\begin{flushright}		CFNUL/96-08
\end{flushright}
\vspace{10pt}

		\begin{center}
{\Large Implications of anomalous gauge boson interactions \\
to the fermion electromagnetic moments}
\\[20pt]{\sc Lu\I s Bento} and {\sc Rui Neves$^\dagger$}\\[10pt]
		\end{center}
\vspace{10pt}
		\begin{center}
{\normalsize \sl Centro de F\I sica Nuclear, Universidade de
Lisboa,\\ 
Av. Prof. Gama Pinto 2, 1699 Lisboa - {\sl codex}, Portugal\\[10pt]
$^\dagger$Department of Mathematical Sciences,\\
	Science Laboratories, University of Durham,\\
	South Road, Durham DH1 3LE, England}

		\end{center}
\vspace{25pt}
		\begin{abstract} 
\normalsize 
\vspace{10pt}
We calculate the electromagnetic form-factors of the  fermions induced
by the anomalous gauge boson interactions contained in the operators
$\ODWW$ and $\ODB$.
The interplay between vertex corrections and gauge boson self-energies
is studied in order to separate the non-universal form-factors.
We apply the same procedure to reannalize previous results regarding
other anomalous gauge boson interactions.
		\end{abstract}
\vspace{40mm}

\begin{flushright}	October 1996 (revised December 96)
\end{flushright}

\vspace{15pt}
\begin{flushleft}
e-mail: LBento@fc.ul.pt  \hspace{0.5mm} and \hspace{0.5mm}
 R.G.M.Neves@durham.ac.uk
\end{flushleft}

\newpage

\section{Introduction}

So far, the standard theory of electroweak interactions is consistent
with
all the available experimental results. The secret of this success
seems
to lie in its gauge symmetry structure, based on the SU(2)xU(1) group
and
served by a Higgs breaking mechanism into $\rm U(1)_{QED}$.
Gauge invariance assures renormalizability, a crucial condition to
evaluate 
and predict higher order corrections, and implies another fundamental
feature concerning the nature of gauge bosons self-interactions.
Unlike the couplings to the matter fields, which require additional
assumptions about the representations of matter (unless some other
principle
is postulated
\footnote{It was showed in ref. \cite{luis} that the observed matter
representations of the gauge groups SU(2) and SU(3) are the only ones
that
satisfy a Principle of Covariance with respect to basis
transformations.}),
the couplings between gauge bosons are completely determined by local
gauge 
invariance. However, no direct precise determination of the $W, Z,
\ga$
self-couplings has been possible so far. 
The finding or elimination of anomalous boson self-interactions will
reveal 
unknown high energy scale dynamics or confirm the very nature of the
weak bosons, as that of the photon is already established. So, a
great part 
of the significance of a machine like LEP2 lies in the possibility of
directly measuring couplings such as $WWZ$ and $WW\ga$.
That is well known and has been studied by many authors
\cite{gg,hag,ruj,mad,lep}.

One may also look at indirect low-energy effects. A lot of work has
been
done focusing either on oblique corrections to 4-fermion amplitudes
\cite{grin,ruj,h2,mad} or on radiative corrections to
the fermion-gauge boson couplings \cite{gm2,mar,bou,bsg,hin,ruj}.
Here, we are particularly interested in non-standard one-loop
corrections to
the electromagnetic moments of the fermions induced by anomalous
$WW\ga$
interactions.
Previous works on that matter were based in one effective Lagrangian
\cite{hag} that is the most general with up to 6-dimensional
tri-linear
operators if the gauge fields are restricted to be transverse
($\partial \cdot W=0=\partial \cdot Z$).
That condition is satisfied if the gauge bosons are on-shell or
coupled to 
massless fermions but is not necessarily true otherwise and
therefore,
there is a potential lack of generality in that mentioned effective
Lagrangian. 
Indeed, we identified \cite{rui} a lot of independent couplings that
vanish 
for transverse bosons but not for the spin zero degrees of freedom,
the
single constraint being electromagnetic gauge invariance. 

Since however, the standard model gauge group has proven to be a good
symmetry, even if in a hidden form, up to scales as high as $\mz$, it
is not 
so natural to expect non-invariant interactions at the scale of New
Physics
$\Lambda$ which is itself above the very Fermi scale of weak
symmetry
breaking.
Moreover, as emphasised by De R\'ujula \etal\ \cite{ruj}, non-
invariant operators produce divergent radiative corrections which,
cut-off 
at the scale $\Lambda$, give rise to deviations from the predictions
of the 
Standard Model such as the $\mw - \mz - \gf$ relation, which no
longer 
decouple in the limit $\Lambda \rarrow \infty$ even if originated
from 
6-dimensional operators suppressed by $1/\Lambda^2$.
The anomalous interactions should then arise from an effective
Lagrangian 
manifestly invariant under local SU(2)xU(1).

In the so-called linear realization, one includes also operators with
the
standard Higgs iso-doublet. A complete set of independent
dimension-6
operators was identified by Buchm\"uller and Wyler \cite{buc}.
Among the $P$ and $CP$ even operators, only seven generate gauge
boson
interactions: $\OF1$, $\OBW$, $\ODW$, $\ODB$ modify the 2-point
Green
functions at tree-level and $\OBW, \ODW, \OB, \OW, \OWWW$, give rise
to
triple gauge boson couplings.
In all cases except one, the $WW\ga$ couplings reduce to
the U(1)-invariant interactions usually cast \cite{hag} in a
phenomenologic Lagrangian parametrized with $\Delta k_\ga$ and
$\lambda_\ga$ ( \eq{LWWg} below ).
Their one-loop effects on the fermion electromagnetic couplings have
been calculated.
The exception is the operator $\ODW$ as it yields interactions that
are not electromagnetic gauge invariant by themselves but are rather
undissociable from anomalous kinetic terms.
That constitutes the primary motivation to study the one-loop
corrections due to $\ODW$ (in fact  $\ODWW$, a linear combination of
$\ODW$ and $\OWWW$) and in particular to determine whether or not it
produces fermion magnetic moments that could be used to set limits on
the anomalous gauge boson interactions.
We also do the calculations for the operator $\ODB$ in view of its
similarity with $\ODW$.

In section 2 we derive the results of the one-loop Feynman diagrams.
In section 3 we study the interplay between vertex corrections and
gauge boson polarization functions including their longitudinal
projections.
It is then shown how to extract the non-universal electromagnetic
couplings.
The same procedure is used to reannalize and briefly overview in
section 4  previous results on the electromagnetic form-factors
induced by triple gauge boson interactions.

\newpage

	\section{$\ODWW$ and $\ODB$ radiative effects}

The effective Lagrangian is a linear combination of SU(2)xU(1)
invariant 
operators,
	\beq
{\cal L}_{\rm eff}={1 \over {\Lambda ^2}}\sum\limits_i {f_i}\,{\cal
O}_i      
\, ,
\label{Leff}
	\neq
that are functions of the Higgs covariant derivatives
and gauge field strength tensors, denoted as (notation of refs.
\cite{mad,h2}),
	\beq
&  & D_\mu \Phi = \left( \partial _\mu +i{g ' \over 2}B_\mu +
i{g \over 2}\, \sigma _i W_\mu ^i \right) \Phi 
	\, , 					\label{DPhi}\\
&  & [D_\mu ,D_\nu ] = \hat B_{\mu \nu }+\hat W_{\mu \nu }=
i{g ' \over 2}B_{\mu \nu }+i{g \over 2}\sigma_i W_{\mu \nu }^i     
	\, . 					\label{DD}
	\neq
The operators $\ODW$ and $\ODB$ are defined as
	\beq
& &\ODW = {\rm Tr} \left\{ \left[ D_\mu ,\hat {W}_{\nu \alpha }
\right]
\left[ D^\mu ,\hat {W}^{\nu \alpha }\right] \right\}
	\, ,                                       \label{ODW}\\
& &\ODB =  -{1 \over 2} {g '}^2
(\partial _\mu B_{\nu \alpha })(\partial ^\mu B^{\nu \alpha })
        \, .				                 \label{ODB}
	\neq
In addition to quadratic terms, $\ODW$ includes also trilinear
couplings, but part of them are already present in
$\OWWW={\rm Tr} \left\{ \hat {W}_{ \alpha \beta }
\hat {W}^{\beta \gamma} {\hat {W}_\gamma }^\alpha  \right\}$
as follows from the identity
	\beq
& &\ODW = -4 \, \OWWW + 2\, \ODWW     \,  ,
						\label{ODW1}\\
& &\ODWW =  {\rm Tr} \left\{ \left[ D_\mu ,\hat {W}^{\mu \alpha }
\right]
\left[ D^\nu ,\hat {W}_{\nu \alpha }\right] \right\}
	\,   .                                          \label{ODW2}
	\neq
Given the fact that $\ODWW$ not only contains all the quadratic terms
of
$\ODW$ and, just like $\ODB$ and unlike $\ODW$, is directly related
through
the equations of motion to the matter currents,
	\beq
& &\ODWW = -{1 \over 2} g^2 J_T^a \cdot J_T^a
	\, ,                                           \label{jj1} \\
& &\ODB = -{g '}^2 J_Y \cdot J_Y
	\, ,                                           \label{jj2}
	\neq
we choose to replace $\ODW$ by $\ODWW$ in the set of linearly
independent operators.
The relation between the coupling constants in this basis
($\ODWW$, $\OWWW$) and in the basis ($\ODW$, $\OWWW$) is
straightforward: $ \fDW=2\, f_{DW}, \fWWW= f_{WWW}-4\, f_{DW} $. 


$\ODB$ and $\ODWW$ generate the following two-point Green functions
($f'_{DB}=2\, f_{DB}$):
	\beq
i\,\Pi ^{\mu \nu }_{*+-} =
-i\mfw\,p^2\left( {p^2 g^{\mu \nu }-p^\mu p^\nu } \right)
							\label{PI+-}
	\neq
for the charged $W$s and
	\beq
i\,\Pi ^{\mu \nu }_{*ab}=
-i\left[ \mfw \, R^3_a R^3_{b} + \mfb \, R^B_a R^B_{b}  \right]
p^2\left( p^2 g^{\mu \nu }-p^\mu p^\nu  \right)
	\, ,						\label{PIab}
	\neq
for the neutral physical particles $\ga$ and $Z$; $R$ is the weak
rotation
matrix from $\ga, Z$ to the $W^3, B$ weak basis.
The  Feynman rule for the anomalous coupling of $W^-_\alpha W^+_\beta
W^3_\mu$
with incoming momenta respectively $x, y, -k$, is
	\beq
-i\,g\,\Ga_* ^{\alpha \beta \mu }= - i\,g\mfw 
& \left\{ \left( x^2\delta _\rho ^\alpha -x^\alpha x_\rho \right)
\Ga_0^{\rho \beta \mu }+
\left( {y^2\delta _\rho ^\beta -y^\beta y_\rho }\right) 
\Ga _0^{\alpha \rho \mu }+  \right. & 		\nonumber\\
& \mbox{ } + \left. \left(k^2 \delta^\mu_\rho - k^\mu k_\rho \right)
\Ga _0^{\alpha \beta \rho } \right\} \, , &
							\label{GDW}
	\neq
	\beq
\Ga _0^{\alpha \beta \mu }=\left( {x^\mu -y^\mu } \right)g^{\alpha
\beta}
-\left( {x^\beta +k^\beta } \right)g^{\alpha \mu }+
\left( {y^\alpha +k^\alpha } \right)g^{\beta \mu }
	\, .						\label{GSM0}
	\neq
For special processes like $W$ pair production at LEP2 this
interaction is not independent from the ones that have been considered
 so far \cite{hag,mad,lep} assembled in the phenomenologic Lagrangian
	\beq
{\cal  L}_{WW\ga}  = -i\,e\, \Delta k_\ga\, W_\mu ^+W_\nu F^{\mu \nu}
-i\, e {\lambda _\ga \over {M_W^2}}W_{\lambda \mu }^+ W^{\mu\nu }
F_\nu ^\lambda
	\; .
\label{LWWg}
	\neq
Indeed, it reduces for transverse bosons ($\partial \cdot W^a=0$) to a
 simple form-factor of the standard model vertex:
	\beq
\Ga_* ^{\alpha \beta \mu } \rarrow
 \mfw (x^2 + y^2 + k^2) \Ga _0^{\alpha \beta \mu }
	\, .						\label{Gstar}
	\neq
However, when considering one-loop effects neither the running with
the momenta nor the longitudinal degrees of freedom can be discarded
in advance.
In addition, as will be seen below, that interaction is not
electromagnetic gauge invariant per se (unlike the ones of \eq{LWWg})
 but only when associated with certain $W$ kinetic terms.
This was the primary motivation to study its one-loop effects.

The standard model $WW\ga$ coupling is just $-i\,\!e\,\!\Ga_0$ in
the
$R_\xi$ gauge but we used the Fujikawa gauge-fixing condition
\cite{Fuji} for the $W$ field namely,
	\beq
{\cal L}_W [{\rm gf}] = - 1/\xi_W \left| \, \partial^\mu W^+_\mu +
i\, e \,\! A^\mu W^+_\mu + i \, g\,\! v \, \xi_W \, \phi^+\! /2 \,
\right| ^2
	\; .						\label{Lgf}
	\neq
It eliminates the trilinear coupling of the photon with the $W$ and
Goldstone
boson $\phi^+$ and in addition, shifts the standard model $WW\ga$
coupling to:
	\beq
\Ga _\sm ^{\alpha \beta \mu } = \Ga _0 ^{\alpha \beta \mu } +
1/\xi _W \left( {x^\alpha g^{\beta \mu }-y^\beta g^{\alpha \mu }}
\right)
	\, .						\label{GSM}
	\neq
The advantage of such $W$ covariant gauge lies in that this $WW\ga$
coupling
and the $W$ full propagator obey a Ward identity:
	\beq
& & -i\,\!e\,\! k_\mu \Ga _\sm ^{\alpha \beta \mu } =
e\left( {G^{-1} \slb y \srb -G^{-1} \slb x\srb } \right)^{\alpha
\beta }
	\, ,						\label{Wid1}\\
& & G^{\alpha \beta } = {i \over {p^2-M_W^2}}\left[ {-g^{\alpha \beta
}+
\left( {1-\xi _W} \right){{p^\alpha p^\beta } \over {p^2-\xi
_WM_W^2}}} \right]
	\, .						\label{GW}
	\neq
Then, since the $\ODWW$ coupling and self-energy satisfy a similar
identity,
	\beq
-i\,\! e \,\! k_\mu \Ga_* ^{\alpha \beta \mu } =
e\left(-i {\Pi _* \slb y \srb +
i\Pi _* \slb x \srb } \right)^{\alpha \beta }_{+-}
	\, ,						\label{Wid2}
	\neq
one obtains an automatic Ward-Takahashi identity for the one-loop
fermion
vertex. Note that the last equation just proves the point that the
above
anomalous $WW\ga$ interaction is not electromagnetic gauge invariant
by itself
 and therefore, cannot be reduced to the couplings of the
phenomenologic Lagrangian of \eq{LWWg}.

We calculated the one-loop corrections to the $f\bar{f} \ga$
vertices
(once for all denoted as $-i \DGg^\mu$) using dimensional
regularization.
There are two kinds of diagrams: the ones where the photon couples to
the
fermion line and the internal gauge bosons carry anomalous
self-energies
and the ones where the photon couples to the charged $W$ (only in the
case of
$\ODWW$) either with anomalous $WW\ga$ coupling or with the standard
one
plus anomalous $W$ self-energy. Keeping only the divergent terms, we
obtained for a on-shell fermion with charge $Q$ and isospin
	\beq
T_3 = t_3 (1-\ga_5)/2 \, , \quad\quad t_3 = \pm 1/2
	\, ,						\label{T3}	
	\neq
the following results after fermion wave function renormalization:
	\beq
& &-i \DGg^\mu = - e \, {g^2 \over 2} \, \mfw \, J
\kk \left[ Q {1-\ga_5 \over 2} - T_3 + 3(1-\xi_W) T_3 \right]
	 ,						\label{DGDW}\\
& &-i \DGg^\mu = - e\, {g '}^2 \, \mfb\, J
\kk {2 \over 3} Q (Q-T_3)^2
	\, .						\label{DGDB}
	\neq
Here, $J$ is the integral in momentum space
	\beq
J=\int {{{d^dp} \over {\left( {2\pi } \right)^d}}}
{1 \over {\left( {p^2-M_W^2} \right)^2}}=
{i \over {16\pi ^2}}\ln {{\Lambda ^2} \over {M_W^2}}
	\; ,						\label{J}
	\neq
where the right-hand side is the result of a cut-off regularization,
the
cut-off scale $\Lambda$ naturally identified with the scale of New
Physics.
The $\xi_W$ dependence of the results is not surprising given the
interplay
between vertex and vacuum polarization functions already present in
the
standard model.

\newpage

	\section{The physical electromagnetic couplings}

It is well known that in non-abelian theories the radiative
corrections to
the fermion-gauge boson couplings are not gauge-invariant per si
neither are
the gauge boson self-energies. Gauge invariant quantities are
obtained as
certain combinations of vacuum polarization and universal vertex
functions.
In the case of the standard model, some of the non-trivial features
can be
reduced to simple corrections of $g$ and 
$g '$ propagated to all gauge boson couplings and masses \cite{ken};
when considering anomalous gauge boson couplings one also finds
additional
mixing between $W^3$ and $B$ in the fermion vertices \cite{h2}.
We present here an extension suitable for massive fermions covering
the interplay between the longitudinal parts of the boson propagators
 and the anapole type of vertex.

In a 4-fermion amplitude at one-loop level, there are vacuum
polarization diagrams
and corrections to the vertices, the latter denoted as
$-i\DG _a^\mu$ for each boson $a=\ga, Z, W^\pm$.
It is convenient to separate the gauge boson propagators and
self-energies 
in their transverse and longitudinal components:
	\beq							
& &G_a^{\mu \nu } = -i\left[ {{{P_T^{\mu \nu }} \over {k^2-M_a^2}}+
{{P_L^{\mu \nu }} \over {k^2/\xi _a - M_a^2}}} \right]
	\, ,	\kern 30 pt  
	\, 						\label{Ga}\\
& &i \,\! \Pi _{ab}^{\mu \nu } = -i\left( P_T^{\mu \nu }\pi _{ab}+
\,P_L^{\mu \nu }\rho _{ab} \right) \,, \kern 50 pt   a, b = \ga, Z,
W^{\pm}
	\, 						\label{PI}
	\neq
where
	\beq
P_T^{\mu \nu } = g^{\mu \nu }-P_L^{\mu \nu } =g^{\mu \nu }-k^\mu
k^\nu /k^2
	\, .						\label{PTPL}
	\neq
By looking at the dependence of the amplitudes on the fermion quantum
numbers,
one comes to the conclusion that a certain kind of one-loop vertices
give the
same results as the vacuum polarization diagrams.
In the case of neutral currents, the most general form of these
universal
couplings is: 
	\beq
\delta \Ga _a^\mu \,[{\rm univ}]
=\ga ^\mu \left( {q\,\Lambda _{\ga a}+q_Z\,\Lambda _{Za}}
 \right)-k^\mu\! {\not{\! k}} \left( {q\,A_{\ga a}+q_Z\, A_{Za}}
\right) \,,
\quad a=\ga, Z
	\, 					\label{dGauniv}
	\neq
where $\Lambda_{ab}$, $A_{ab}$ are flavour independent functions of
$k^2$ and
$q\!=\!Q\,\!e$, $q_Z$ are the electric and $Z^0$ fermion charges:
	\beq
q_Z =\left( {T_3- Q \kern 1pt \sin^2 \theta_W } \right) \sqrt
{g^2+g'^2}
	\, .						\label{qz}
	\neq
Of course, one could write the universal couplings in terms of $Q$
and $T_3$
as well, but the above formulation makes it particularly easy to show
that the
sum of vertex and vacuum polarization diagrams
remains invariant if one replaces the self-energies and couplings
with:
	\beq
& &\pi '_{ab}  =\pi _{ab}-( {k^2-M_a^2} )\Lambda _{ab}-
( {k^2-M_b^2} )\Lambda _{ba}
	\,,						\label{pi1}\\
& &\rho '_{ab} =\rho _{ab}-( {k^2/\xi _a-M_a^2} )
( {\Lambda _{ab}-k^2 A_{ab}})-
( {k^2/\xi _b -M_b^2} )( {\Lambda _{ba}-k^2 A_{ba}} )
	 ,						\label{rho1}\\
& &{\DG '}_a^\mu  =\DG _a^\mu -
\ga ^\mu \left( {q\,\Lambda _{\ga a}+q_Z\,\Lambda _{Za}}
 \right)-k^\mu\! {\not{\! k}} \left( {q\,A_{\ga a}+q_Z\, A_{Za}}
\right)
	\, .						\label{DG1}
	\neq
In this way one can obtain gauge-invariant quantities\footnote
{Eq. (\ref{pi1}) is in agreement with ref.\ \cite{h2} where the
same relations were derived for particular $\Lambda$ functions.}.

We define the electromagnetic field as the one that couples
universally with the electric charge exclusively and obeys the
Maxwell equations of motion which implies a zero photon mass
and a dynamical decoupling from the $Z^0$ field.
The first condition is realized by the cancellation of the universal
part of
$\DGg ^\mu$ proportional to $q_Z$ by the $\Lambda_{Z\ga}$ and
$A_{Z\ga}$
terms. Hence, the remaining universal component of the
electromagnetic
coupling reduces to a running electric charge unit $e(k^2)$:
	\beq
{\DG '}_\ga^\mu \,[{\rm univ}] = Q \,\Delta e(k^2) \, \ga ^\mu
	\,  .					\label{DGguniv}
	\neq
The second condition is realized by chosing the functions
$\Lambda_{ab}$ so
as to annihilate the real part of the renormalized polarization
functions:
	\beq
\Re \left(	\begin{array}{cc}
\pi '_{\ga \ga}(k^2) & \pi '_{\ga Z}(k^2) \\
 & \\
\pi '_{\ga Z}(k^2)  &  \pi '_{ZZ}(k^2)-\pi '_{ZZ}(M_Z^2)
\end{array}	\right) \equiv 0
	\, .						\label{Rpi}
	\neq
As a result, the renormalized transverse inverse propagator is just
given by
	\beq
\left(	\begin{array}{cc}
k^2 + \Im  \left\{ \pi _{\ga \ga} \right\} &
\Im \left\{ \pi _{\ga Z} \right\} \\
 & \\
\Im \left\{ \pi _{\ga Z} \right\}	&
k^2 -M^2_Z + \Im  \left\{ \pi _{Z Z} \right\}
\end{array}	\right)
	\, ,						\label{Rtip}
	\neq
where $\Im$ stands for the imaginary part: the $\ga - Z$ decoupling
is manifest.
Finally, $A_{\ga\ga}$, $A_{\ga Z}$ and the longitudinal polarization
functions
$\rho_{\ga \ga}$, $\rho_{\ga Z}$ are immaterial for on-shell fermion
amplitudes, as they give contributions proportional to the operator
$q\,\kks$
that vanishes in that case.

\pagebreak

The solution of the above set of conditions is then:
	\beq
& &\Lambda _{\ga \ga }=
\,\Re \left\{ {\pi _{\ga \ga }(k^2)} \right\}/2k^2
	\, ,						\label{Lgg}\\
& &\Lambda _{\ga Z}=\, \Re \left\{ {\pi _{\ga Z}(k^2)-
( {k^2-M_Z^2} )\Lambda _{Z\ga }} \right\}/k^2
	\, ,						\label{LgZ}\\
& &\Lambda _{ZZ}=\,\Re \left\{ {\pi _{ZZ}(k^2)-\pi _{ZZ}(M_Z^2)}
\right\}
 /2( {k^2-M_Z^2} )
	\, .						\label{LZZ}
	\neq
When these expressions are substituted in \eq{DG1} one obtains for
the neutral current amplitudes what is called improved Born
approximation \cite{ken,con}: it means that the sum of the amplitudes
with boson self-energies and vertex radiative corrections is written
with expressions where the gauge boson propagators are the free
particle propagators (except for imaginary parts, see \eq{Rtip}) and
the radiative corrections are collected in the new
{\em gauge-invariant} vertices given by \eq{DG1}.
They contain both flavour dependent and universal form-factors.
The latter, once added to the tree-level vertices, can be expressed
as running coupling constants.
The boson self-energies contribute only to the universal vertices as
follows: $\Lambda_{\ga\ga}$ contributes to the running electric
charge unit and $\Lambda_{Z Z}, \Lambda_{\ga Z}$ to the running
coefficients of $T_3$ and $Q$, in the $Z^0$ coupling (\eq{qz}).

In what regards the effects induced by $\ODWW$ and $\ODB$ it is not
difficult to isolate from \eqs{DGDW}{DGDB} the flavour dependent
electromagnetic coupling as
	\beq
\DGg^\mu [f] = Q \kk
\left[ -\beta \,\mfw \,\ga _5 + {8 \over 3}\beta^\prime \, \mfb
( Q-T_3 )^2 \right] \ln {{\Lambda ^2} \over {M_W^2}}
	\, ,						\label{DGf}
	\neq
where
	\beq
\beta = {{g^2e} \over {64\pi ^2}} 	\, , \quad \quad \quad
\beta  ' = {{ {g '}^2 e} \over {64\pi ^2}}
	\, .						\label{beta}
	\neq
This expression is clearly gauge independent and does not receive
contributions from the boson self-energies.
The remaining terms can be put in the form of \eq{dGauniv} and have
to be summed to the universal contributions arising from the boson
self-energies as explained before.
As far as the electromagnetic interaction is concerned the overall
result is a running $\alpha_{\rm QED}$ but such kind of contribution
already appears in the tree-level self-energy specified by \eq{PIab}
yielding
	\beq
\Delta e(k^2) = - e\, \pi_{\ga \ga}/2k^2 =
 - e^3 (f'_{DW} + f'_{DB} ) k^2 /2\Lambda^2
	\, .						\label{De}
	\neq
This is certainly the leading term of the universal electromagnetic
coupling and for that reason is not worth to calculate the one-loop
boson self-energies.
Furthermore, that and other oblique corrections have been annalized
by several authors \cite{ruj,grin,h2,mad} and will not be further
studied here.
Our primary interest were the non-universal flavour dependent
electromagnetic form-factors.
The result shown in \eq{DGf} contains no magnetic moment term and
comprises a charge radius (CR) and one anapole moment (AM) whose
values depend on the charge and isospin quantum numbers of each
particle.
But, since they are proportional to the electric charge both vanish
in the neutrino case.

In processes at very low energies, the CR and AM contributions are not
dynamically different from other local interactions such as the ones
mediated by $Z^0$. That fact gives the opportunity for adopting
different
definitions of CR and AM.
Although such a discussion is out of the scope of this work, we just
add that
within the presentation given in this section, the CR and AM arise as
the
local interaction couplings that survive in four fermion amplitudes
if the
source of the
electromagnetic and $Z^0$ fields has a zero $q_Z$ charge (\eq{qz}).
That is approximately true if the source (target) is a medium made
of
unpolarized electrons and/or protons.
But in the work of G\'ongora and Stuart \cite{robin} for instance,
the CR and
AM are the couplings that survive if the source has a zero $t_3$
isospin
component.
One should keep in mind however, that what matters is the total
amplitude
which may comprise other local interaction contributions such as box
diagrams, or even charged current interactions in the case of
elastic
scattering. Actually, both of them are produced by the operators
$\ODWW$,
$\ODB$ at one-loop level.

\newpage
\section{Overview and conclusions}
	
Boudjema \etal\ \cite{bou} calculated (in the unitary gauge) the
fermion
form-factors generated radiatively by the anomalous $WW\ga$
interactions
specified by the Lagrangian of \eq{LWWg}.
We are now in position to recognize that the couplings they obtained
are just the universal kind of vertices with only one exception.
Keeping only the divergent terms, one has for a fermion with mass $m$
and isospin $t_3 = \pm 1/2$:
	\beq
& &\DGg ^\mu =a \,\kk T_3
- i\,\Delta \mu \, \sigma ^{\mu \nu }k_\nu
	\, ,					\label{DGkl}\\
& &a = \beta\, {{\Delta k_\ga } \over {M_W^2}}{{\Lambda ^2} \over
{M_W^2}}
+ 4\, \beta\, {{\lambda _\ga } \over {M_W^2}}\ln {{\Lambda ^2} \over
{M_W^2}}
	\, ,						\label{a}
	\neq
where the anomalous magnetic moment takes the value (also calculated
in refs. \cite{gm2}):
	\beq
\Delta \mu = 2\,m\,t_3 \,\beta\,
{{\Delta k_\ga } \over {M_W^2}}\ln {{\Lambda ^2} \over {M_W^2}}
	\, .						\label{mu}
	\neq
It is clear that the $a$ term is a particular case of the universal
interactions identified in \eq{dGauniv}.

More recently, Hagiwara \etal\ \cite{h2} calculated the one-loop
fermion gauge
couplings arising from the SU(2)xU(1) invariant operators $\OWWW,
\OW, \OB$. 
They restricted to the chiral conserving vector and axial-vector
form-factors
in the limit of zero fermion masses and found that $\OB$ does not
contribute
and only $\OWWW$ corrects the electromagnetic coupling.
For a finite fermion mass the result is
	\beq
\DGg ^\mu = 6\, \beta\,
{{f_{WWW}\kern 1pt g^2} \over {\Lambda ^2}} \ln {{\Lambda ^2} \over
{M_W^2}}
\kk T_3
	\, .						\label{DGWWW}
	\neq
In view of the relations \cite{h2} between the parameters of the
phenomenologic and effective Lagrangians (\eqs{Leff}{LWWg}) namely,
	\beq
& &\frac{\Delta k_\ga}{M_W^2} =\frac{f_B+f_W}{2\Lambda ^2}
	\, ,					\label{deltak}\\
& &\frac{\lambda _\ga}{M_W^2} =3\, \frac{f_{WWW}}{2\Lambda ^2} \, g^2
	\, ,					\label{lambda}
	\neq
there is agreement in the $\lambda_\ga - f_{WWW}$ case (the $WW\ga$
operators
are exactly the same), but the two results seem in conflict in what
regards
the charge radius and anapole moment proportional to $\Delta k_\ga$.
That is not necessarily true because those are just the kind of
universal
couplings (\eq{dGauniv}) not expected
to be independent from the gauge-fixing condition.
Boudjema \etal\ worked in the unitary gauge and did not calculate the
vacuum
polarization functions whereas the 't-Hooft-Feynman gauge was used
in
ref. \cite{h2}.

In conclusion, the non-universal electromagnetic form-factors
contained in \eqs{DGkl}{DGWWW} reduce just to a magnetic moment whose
value is only significant (cf. \eq{mu}), given the available
experimental data, in the case of the muon.
The associated $WW\ga$ coupling proportional to $\Delta k_\ga$ is
produced either in the operators $\OW$ and $\OB$ or in $\OBW$
\cite{ruj,h2}.
In turn, the operator $\OWWW$ only yields oblique corrections that
simply renormalize the coefficients of $\ODWW$ and $\ODB$ \cite{h2}.
The analysis of the low energy constraints on the oblique corrections
performed by Hagiwara \etal\ \cite{h2} and updated in \cite{mad}
gives upper limits of the order of 2 and 40 to the absolute values of
$\fDW$ and $f'_{DB}$ respectively for a scale $\Lambda=1\, {\rm TeV}$.
The operators $\ODWW, \ODB$ do induce flavour dependent charge radius
and anapole moments (\eq{DGf}) but not a magnetic moment.
Therefore, although their quantum number structure is different from
the tree-level oblique corrections, the magnitude is
further suppressed by a factor $\beta/e$ of the order of $10^{-4}$
and cannot be used to give new limits on the anomalous interactions.
Finally, some anomalous $WW\ga$ interactions also have effects on the
flavour changing $b \rarrow s \,\! \ga$ transition \cite{bsg,hin},
but we checked that this is not the case of $\ODWW$ and $\ODB$.

\vspace{60pt}
\begin{flushleft}
{\bf \large Acknowledgements}

R. G. was supported in part by the grants BM/2190/91, BD/2828/93
and L. B. by the projects PCERN/P/FIS/118/94, CERN/P/FAE/1050/95.
\end{flushleft}

\end{document}